\documentclass[journal]{IEEEtran}
\usepackage{graphicx}
\usepackage{textcomp}
\usepackage{nth}

\begin{document}
\title{FACT --\\Operation of the First G-APD Cherenkov Telescope}

\newcommand{\ethz}{$^1$}
\newcommand{\tudo}{$^2$}
\newcommand{\unige}{$^3$}
\newcommand{\uniw}{$^4$}
\newcommand{\epfl}{$^5$}

\newcommand{\uniz}{$^{a}$}
\newcommand{\kynu}{$^{b}$}
\newcommand{\mpim}{$^{c}$}
\newcommand{\tum}{$^{d}$}

\author{
T.~Bretz\ethz$^{,*}$,\\
A.~Biland\ethz,
J.~Bu\ss\tudo,
V.~Commichau\ethz,
L.~Djambazov\ethz,
D.~Dorner\uniw,
S.~Einecke\tudo,
D.~Eisenacher\uniw,
J.~Freiwald\tudo,
O.~Grimm\ethz,
H.~von Gunten\ethz,
C.~Haller\ethz,
C.~Hempfling\uniw,
D.~Hildebrand\ethz,
G.~Hughes\ethz,
U.~Horisberger\ethz,
M.~L.~Knoetig\ethz,
T.~Kr\"ahenb\"uhl\ethz,
W.~Lustermann\ethz,
E.~Lyard\unige,
K.~Mannheim\uniw,
K.~Meier\uniw,
S.~Mueller\tudo,
D.~Neise\tudo,
A.-K.~Overkemping\tudo,
A.~Paravac\uniw,
F.~Pauss\ethz,
W.~Rhode\tudo,
U.~R\"oser\ethz,
J.-P.~Stucki\ethz,
T.~Steinbring\uniw,
F.~Temme\tudo,
J.~Thaele\tudo,
P.~Vogler\ethz,
R.~Walter\unige,
Q.~Weitzel\ethz\\
(FACT Collaboration)\\[1ex]
$^{*}${\em Corresponding author: Thomas Bretz (tbretz@phys.ethz.ch)}\\[1.5em]
\thanks{\ethz ETH Zurich, Switzerland --
   Institute for Particle Physics, Schafmattstr.~20, 8093 Zurich}
\thanks{\tudo Technische Universit\"at Dortmund, Germany --
   Experimental Physics 5, Otto-Hahn-Str.~4, 44221 Dortmund}
\thanks{\unige University of Geneva, Switzerland --
   ISDC, Chemin d'Ecogia~16, 1290 Versoix --
   DPNC, Quai Ernest-Ansermet 24, 1211 Geneva}
\thanks{\uniw Universit\"at W\"urzburg, Germany --
   Institute for Theoretical Physics and Astrophysics,
   Emil-Fischer-Str.~31, 97074 W\"urzburg}
\thanks{\epfl EPFL, Switzerland --
   Laboratory for High Energy Physics, 1015 Lausanne}
\thanks{\uniz Also at: University of Zurich, Physik-Institut, 8057 Zurich,
   Switzerland}
\thanks{}
\thanks{The important contributions from ETH Zurich grants ETH-10.08-2 and
ETH-27.12-1 as well as the funding by the German BMBF (Verbundforschung
Astro- und Astroteilchenphysik) are gratefully acknowledged. We thank
the Instituto de Astrofisica de Canarias allowing us to operate the
telescope at the Observatorio Roque de los Muchachos in La Palma, the
Max-Planck-Institut f\"ur Physik for providing us with the mount of the
former HEGRA CT\,3 telescope, and the MAGIC collaboration for their
support. We also thank the group of Marinella Tose from the College of
Engineering and Technology at Western Mindanao State University,
Philippines, for providing us with the scheduling web-interface.}
}

\maketitle
\pagestyle{empty}
\thispagestyle{empty}

\begin{abstract}
Since more than two years, the First G-APD Cherenkov Telescope (FACT)
is operating successfully at the Canary Island of La Palma. Apart from
its purpose to serve as a monitoring facility for the brightest TeV
blazars, it was built as a major step to establish solid state photon
counters as detectors in Cherenkov astronomy. 

The camera of the First G-APD Cherenkov Telesope comprises 1440
Geiger-mode avalanche photo diodes (G-APD aka. MPPC or SiPM) for photon
detection. Since properties as the gain of G-APDs depend on
temperature and the applied voltage, a real-time feedback system has
been developed and implemented. To correct for the change introduced by
temperature, several sensors have been placed close to the photon
detectors. Their read out is used to calculate a corresponding voltage
offset. In addition to temperature changes, changing current introduces
a voltage drop in the supporting resistor network. To correct changes
in the voltage drop introduced by varying photon flux from the
night-sky background, the current is measured and the voltage drop
calculated.  To check the stability of the G-APD properties, dark count
spectra with high statistics have been taken under different
environmental conditions and been evaluated.

The maximum data rate delivered by the camera is about 240\,MB/s. The
recorded data, which can exceed 1\,TB in a moonless night, is compressed
in real-time with a proprietary loss-less algorithm. The performance
is better than gzip by almost a factor of two in compression ratio and
speed. In total, two to three CPU cores are needed for data taking.
In parallel, a quick-look analysis of the recently recorded data is
executed on a second machine. Its result is publicly available within a
few minutes after the data were taken.

The telescope is the first Cherenkov telescope which is operated 
completely remote. Its data taking is fully automatic, although
interventions by the remote shifter are still necessary in
case of events like sudden weather changes.


\end{abstract}


\section{Introduction}

\begin{figure}[htb]
\centering
\includegraphics[width=.5\textwidth]{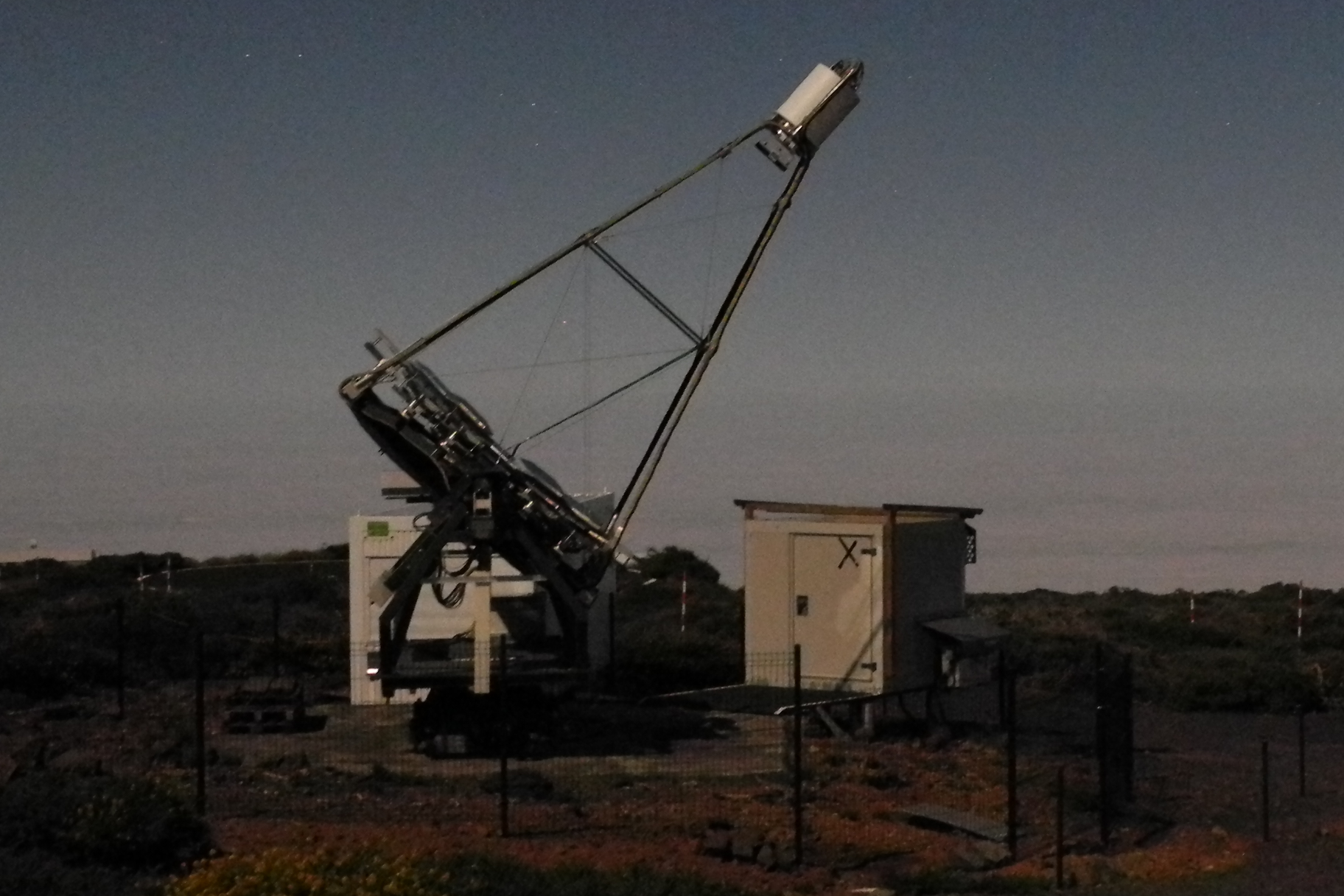}
\caption{Image of the FACT telescope during observations at night. The reflector
is seen from the side, the lids of the open camera are visible. Behind the
telescope the container with the power supplies and control PCs is visible.}
\label{fig:fact}
\end{figure}

For the first time, a Cherenkov telescope is operated using Geiger-mode
avalanche photo diodes (G-APD aka. SiPM) as photo detectors. It is also
the first time that a Cherenkov telescope is operated remotely. The
First G-APD Cherenkov Telescope (FACT) as shown in Fig.~\ref{fig:fact}
has been operated for three years and situated at the Observatorio Roque de
los Muchachos at the Canary island of La Palma (Spain). First results
were published in~\cite{bib:gamma,bib:icrc}.

\begin{figure}[htb]
\centering
\includegraphics[width=.5\textwidth]{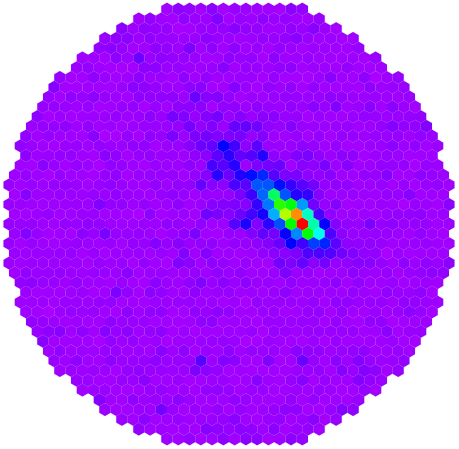}
\caption{Example of a recorded shower image. The color scale (from violet to red) 
is linear and corresponds to the signal amplitude recorded in each pixel.}
\label{fig:shower}
\end{figure}

The imaging air-Cherenkov technique is an indirect measurement
technique, which is necessary as gamma-rays do not penetrate the
Earth's atmosphere down to ground. Instead, Cherenkov telescopes record
light flashes of the Cherenkov light emitted by the particle cascade
induced by a primary particle hitting the Earth's atmosphere. Using the
recorded light intensity and distribution allows to reconstruct energy
and direction of the primary particle as well as its type. An example
of a recorded shower image is shown in Fig.~\ref{fig:shower}.

To detect these faint light flashes which last only several
nano-seconds, a fast and very sensitive camera is necessary. In the
past, such cameras have used photo multiplier tubes for photo
detection. The focal plane is equipped with 1440 sensors each
readout individually. Each sensor is glued to a solid light guide to
increase the light collection area of the sensor. The field-of view of
each sensor is 0.11\textdegree{} providing a total field-of-view of
4.5\textdegree{}.

The response of silicon based photo sensors depends on the ambient
temperature and the applied voltage. Therefore, a feedback loop
is necessary which corrects for the temperature effect and the
voltage drop induced by changing ambient light levels.

For data acquisition, the Domino Ring Sampler DRS\,4 is used. Under 
standard data taking conditions, the waveform is sampled with
2\,Gsamples and a readout window of 150\,ns. The data is transferred
via Ethernet to a standard PC. A proprietary loss-less compression
algorithm compresses the data in real time while it is written to disk.

Apart from the data acquisition software, many other systems have to be
controlled and read out. A common software framework for all programs
ensures easy maintenance and simplifies the implementation. To
control the system centrally, a JavaScript interpreter is used. The use
of JavaScript allows easy scripting while forcing some simplicity for
the scripts.

The common framework, the application of modern programming techniques
and the simplicity of the implemented programs ensure a stable
operation.

The observation schedule is stored in a central database and can 
be edited from a web interface. This allows full automation of the
system. While some steps during data taking as start-up and shutdown or
the reaction on significant worsening of weather conditions are still
partially manual for security reasons, data taking itself is
fully automatic.

A web based monitoring and control allows full control over the
system operated under standard conditions. In other cases, each program
can be controlled individually from a console interface.

An analysis running in parallel just after data recording ensures the 
availability of flux estimates a few minutes after data taking.

The following will give a short introduction to the feedback system
which has been discussed in more details in~\cite{bib:feedback} together
with the results of the measurements. It will give a more detailed
description of the implementation of the data acquisition software, the
data compression, the slow control software and the remote control
software.

More technical details about the hardware and software are provided
in~\cite{bib:design}.

\section{Feedback}

The response of silicon based photo sensors depends on their
temperature and the applied voltage. To correct the temperature effect,
their temperature can be measured and the applied voltage adapted
accordingly. The applied voltage can change due to the increased
voltage drop at the supporting resistors when the ambient light level
is changing. To correct this effect some kind of feedback is necessary.

\begin{figure}[htb]
\centering
\includegraphics[width=.49\textwidth,clip,trim=1cm 2cm 0.5cm 6cm]{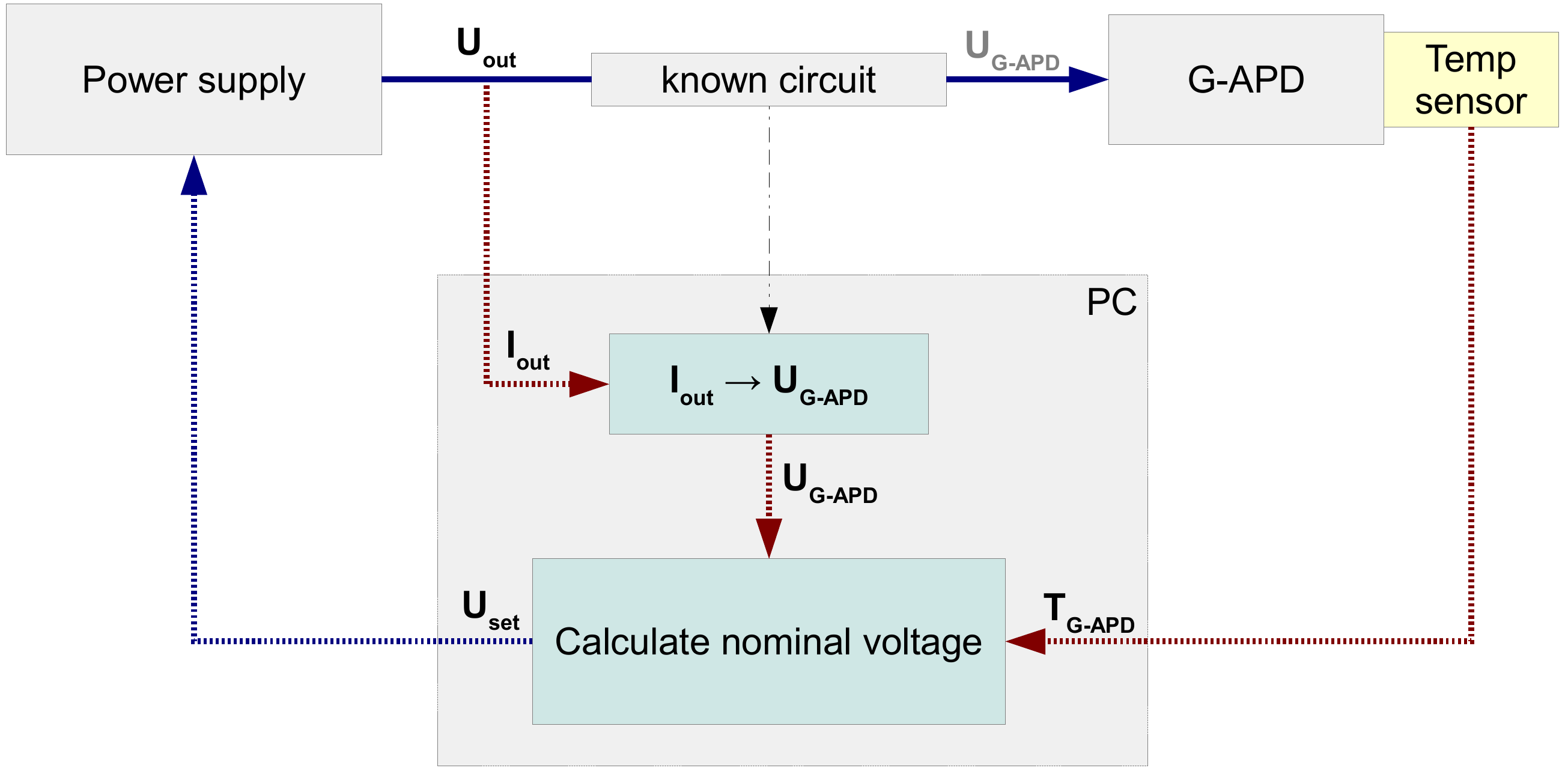}
\caption{Sketch of the feedback loop. The power supply provides the bias
power for the G-APDs through a resistor network. From the measured current in
each channel, the voltage drop and hence the voltage applied to the sensors
is calculated. Together with the reading from the close-by temperature sensors,
the nominal voltage is derived and sent back to the power supply.}
\label{fig:feedback}
\end{figure}

In the FACT camera, the temperature of the focal plane is measured at
28 positions and interpolated. To close the feedback loop for the voltage
control, the current is measured about once a second for each voltage
channel and the corresponding voltage drop calculated. The output
voltage of the power supply is then adapted for each voltage channel
accordingly. A sketch of the loop is shown in Fig.~\ref{fig:feedback}.

That this keeps the sensor properties stable within the limit of the
power supply has been demonstrated in~\cite{bib:feedback}. For that,
dark counts have been recorded and analyzed and the corresponding dark
count spectra compiled. The fact that they can be super-positioned
although taken at different temperatures, shows impressively the
precision of the system.

\section{Data acquisition}

For the total number of 1440 readout channels, 160 Domino Ring Sampling
chips (DRS\,4) are used mounted on 40 readout boards.  Each readout
board has an Ethernet port and is connected to a switch from which four
Ethernet cables are connected to an Ethernet card in a PC. The maximum
throughput achieved is around 240\,MB/s which corresponds to the
maximum data rate the used Ethernet chips can provide. Under normal
data taking conditions, sampling 150\,ns with 2\,Gsamples and trigger
rates below 100\,Hz, the data rate is well below that limit. For some
calibration data, the system is operated close to its throughput limit.
As protocol, the TCP/IP protocol is used, because it ensures that packages
do not get lost and that they arrive in the correct order. The data
acquisition software is implemented in C++11 and is using the
boost-libraries. 

There are several ways to read data from several Ethernet ports (poll,
epoll, etc.), but tests have shown that the most efficient way is a
continuous non-blocking read from a socket (recv). There is no real need
to run more than a single thread reading from the 40 open sockets and
the additional need for synchronization would significantly increase
complexity.

The maximum size of data which can arrive from each board per event is
pre-allocated and used as a static buffer. Once the first header of any
of the 40 boards of an event has arrived, the memory for the whole
event is allocated. Since allocation and de-allocation can become a
time consuming operation when large memory chunks are handled or memory
becomes fragmented, the memory is allocated in chunks corresponding to
the maximum event size and never freed. Each free operation is just
putting the corresponding pointer back on a stack from where it is
retrieved if new memory is requested. Physical allocation takes only
place when not enough memory is allocated or available. To avoid
conditions in which the system runs out of memory, the maximum amount
of memory which can be allocated is limited. If this limit is reached,
no further read operation takes place until free memory becomes
available again.

Pointers to the event objects are stored and handled by shared pointers
(std::shared\_ptr), so that memory is kept allocated until all
references to the event have vanished and is automatically freed once
no reference are left. This ensures efficient memory
handling and intrinsically avoids segmentation faults. To keep a list
of incomplete events, a double-linked list (std::list) is used keeping
a constant access time for adding and removing entries.

Once an event has been compiled from all 40 boards and is completely
available in memory, its shared pointer is propagated to a Queue. A
Queue is a thread which is processing the data in a queue with a
pre-defined processing function. Using a conditional mutex the thread
can be sleeping until it gets woken up by a signal which is
automatically emitted when a new entry is added to the queue. It will
only fall asleep again when there are no further events in the queue.
In this way, CPU time is not wasted for waiting operations keeping an
efficient handling of the events. If an event was successfully
processed, the shared pointer is either released or handed over to a
new queue. This concepts ensures a self consistent handling of the
memory with a high flexibility and also allows an easy implementation
of new algorithm processing the data. Only access to the stack of
pointers needs to be protected against race conditions by mutex
semaphores.

Currently several queues are available and executed
sequentially. The first queue only tests the integrity of the events.
In the secondary queue, events are written to disk. Whenever the
run-number in the sequence of events has changed, a new file is opened
automatically.

In parallel, the event is posted to a processing queue, where some
calibration is applied which is  necessary to calibrate the features of
the DRS\,4, so that a first analysis of each event is available.
Within an interval of three seconds, the event with the highest signal
is selected and available for display in a graphical interface to have
an easy way to monitor the system. 

Other parallel queues propagate header information, counters and other
meta data extracted from the events to the slow control system.

No further analysis is applied yet, but the implementation of a
software trigger within this concept should be trivial.

\section{Data compression}

The data is written to FITS files. The FITS format is a format widely
spread in astronomy. Its main purpose is writing of tables with 
a predefined row-size. The data format is described in a well-defined
header which makes the format self-consistent. If the binary data
written to the tables is compiled in memory, it can be written as a
direct stream to the files as one stream which makes it an efficient
data format for continuous data streams. Decompilation of the data can
be done in a similar manner once data is read from the file. Also here
the advantage is a very efficient access and the possibility to read
the data as a single stream, interesting for example if compression
algorithms like gzip are applied. Using the z-library (libz), data can
be decompressed during a normal read operation so that compressed files
can be read directly. However, the compression ratio of this library
(or gzip) is not optimal for the raw data, therefore, a
proprietary algorithm was implemented.

In a first step, the offsets calculated for each capacitor in all
DRS chips is subtracted as an integer offset. This operation is easily
reversible and significantly decreases the amount of random noise in the
sampled waveforms. In a second step, the data is smoothed to keep the
number of necessary tokens low. Therefore, from each sample, the
average of the two previous samples is subtracted which is a good
estimate of the next sample. The advantage is that this operation is
fully reversible. The averaging of two samples decreases the amount of
noise in the subtracted value significantly. As now 16-bit values 
are available of which in almost all cases the hi-byte is zero, a
simple compression algorithm like a byte-wise Huffman encoding  would
still need at least one bit to encode the hi-byte. Therefore, a
word-wise Huffman encoding is applied.

The implemented compression algorithm is significantly faster than
applying a z-lib compression and compresses the data by almost a factor
of two better. If compressed event by event, the algorithm can easily
be parallelized. This allows to easily run the compression in real
time during data taking while saving a significant amount of disk space
and lowering the I/O demands on the system. During standard data
taking, typically less than one core is enough for data compression. In
case of maximum trigger rates, about three cores are needed.

\section{Slow control}

The slow control system is based on a common framework and
communication via the Distributed Information Management System
(DIM,~\cite{bib:dim}). The DIM-network allows to send commands to servers as
well as updating so-called services, which can be considered a
broadcast of a piece of information to all clients. The configuration
of the whole system is stored in a database and re-loaded from the 
database at each start of a program. Changing of the setup during runtime
is intentionally not foreseen to avoid easy change of the setup
during data taking. A web interface is available to edit the setup.
A history of all changes is avilable and a web interface to
to compare configurations from different dates and times.

To ensure stable operation of the system, generally memory is never
allocated manually, instead flexible data members like std::vector are
used. Having a central event queue in each program avoids the need for
any synchronization which significantly reduces complexity. Events can
either be triggered by the user via commands arriving through the
DIM-network, console input processed by the readline library or service
updates. The central event queue ensures sequential processing which
significantly simplifies the code dealing with asynchronous inputs.

Since each DIM command comes with the format of its parameters, a
command entered in the console interface can be directly converted into
a command sent over the network or to the local client. This avoids the
need to implement a console based command interface and enables the
easy implementation of a central command interface. To simplify the
interpretation of the command arguments, each server automatically
offers a additional description service which distributes the
description of the command parameters as well. A similar service is
available for the DIM services, so that for each value in a DIM service
a name, a unit and a description is available.

A central data logger is logging all service updates in the DIM
network, so that all values which are available from each server are
automatically logged to files. Human readable logging is also 
broadcasted as DIM service and hence logged by the central data logger
as well.

Since all information is available as DIM service and all actions can
be executed via commands, it is very simple to implement a graphical
user interface. Since many user interfaces (as Qt\,4) implement central
event queues, received data can be propagated to such a queue and
displayed in graphical elements. Buttons and other controls can be
connected to DIM commands. This allows a very simple implementation of
a graphical interface just using a GUI builder as the qt4-builder.

For the implementation of a central control, a JavaScript interpreter has
been chosen which allows to enhance the language with callbacks to
C-functions. This has two advantages: The simplicity of JavaScript
forces the implementation of higher level functionality such as
histogramming somewhere else which keeps the central scripts simple and
therefore easy to maintain. The second advantage is that the JavaScript
sandboxes can be fully integrated and controlled by the existing
framework. The fact that there is full control over the interactions of
the script from outside ensures that the integrity of the system can be
maintained at any time improving the stability of the system
significantly.

Having this framework available, implementation of communication with
additional hardware components or new clients is reduced to the setup
of the user commands, the implementation of the action based on the
commands and the implementation of handling data received from 
possible hardware components.

Since most of the intelligence is based in the framework and therefore 
implementation of the final programs remains simple, a high level of
stability is ensured important for any remotely and automatically
operated instrument. This is demonstrated by a typical data taking
efficiency of more than 90\% per night, mainly limited by the time
needed to re-position the telescope, as shown in~\cite{bib:IEEENSS}.

\section{Remote control}

\begin{figure}[htb]
\centering
\includegraphics[width=.49\textwidth]{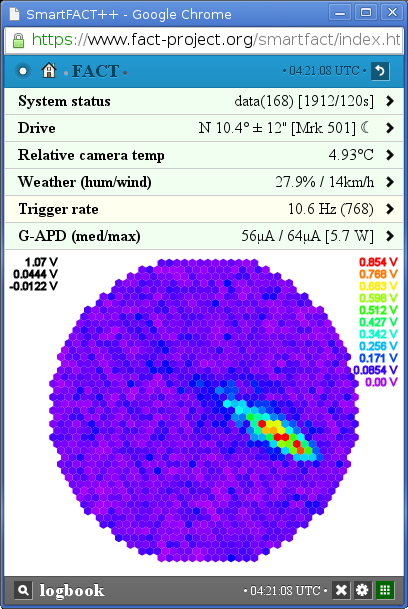}
\caption{Example of the main page of the web interface for monitoring
and control during data taking.}
\label{fig:smartfact}
\end{figure}

For remote control, a web interface~\cite{bib:smartfact}
(Fig.~\ref{fig:smartfact}) has been implemented. Its back-end is a
client which subscribes to all DIM services in the DIM network, similar
to the central data logger. The data received is then processed and the
contents of small ASCII files are updated accordingly. The ASCII files
contain the raw values, or an encoded history of values or a list of
values corresponding to the channels in the camera. 

To keep the traffic low, the page layout and the displayed data
is transferred separately so that the continuous data stream is reduced
to the raw information which is really updated. By loading the
page, mainly a small JavaScript is retrieved. After the JavaScript has
loaded the layout of the requested page, the page is compiled
interactively in the browser and the data loaded and updated in
regular intervals. This allows transfer rates of less than 1\,kB/s
which is low enough that even monitoring from a mobile phone with a
slow internet connection is possible. At the same time, this also
limits the possibility for page layouts which unifies the control
and therefore ensures an easy understanding of the functionality.

If a page is marked as a command-page, a button is displayed allowing to
submit form-data or just a command to an interpreting php. The
php then calls a program which injects the command into the DIM network
and therefore propagates it to the corresponding server. This allows
direct control of all functions needed during
standard data taking and, in turn, avoids the need to be logged in on
the machines running the system and thus keeps uncontrolled user
interference to a minimum, a requirement for each stable system.

\section{Automation}

To automate operation, a central JavaScript takes care of the operation
of the telescope. Its main purpose is to convert a schedule stored in a
database into actions. In addition, it ensures to bring the
whole system back into a well controlled condition once started, so
that in the rare cases of failure a restart is enough. Typical
actions during standard data taking are opening or closing the lid, 
switching the bias voltage of the sensors on or off, re-position
the telescope and start data recording. At this high level some actions
can be executed synchronously to gain observation time.

The schedule can be coded as JSON object and entered into the database
via a locally executed JavaScript or from a web-page. The schedule 
stores a list of observations and tasks. Each observation has a fixed
start time and can contain several tasks which are executed
sequentially. If the start time for the next observation has passed,
the current run or action is finished and the next observation started
right after. To be flexible, a configuration formatted as JSON string
can be added to each task. This allows the implementation of many
possible configurations.

In this system, only the last task in an observation can be of an
undefined length. While the previous tasks have to be limited in time,
the last task, e.g. standard data taking, might be executed until the
next scheduled observation. Assuming a potential problem in the
data taking procedure or a newly discovered error, a restart of the
program could easily re-execute the first task of an observation over
and over again. Therefore, if the start time has already passed,
only the last task of the current observation is executed. If the
script is started before the scheduled time, all tasks are executed
normally. To avoid that after restart, the last scheduled task is
executed again, e.g the operator wants to wait for better weather
conditions, an idle-task can be scheduled.

\section{Conclusions}

The First G-APD Cherenkov Telescope (FACT) is the first Cherenkov
telescope which is operating with silicon photo sensors and which is
fully remote controlled. With the implementation of a feedback system,
it has been proven that the respose of those sensors can be kept stable
within the limits of the power supply, and effects from changing light
level and ambient temperatures can be corrected. 

A well designed data
acquisition software ensures stable operation of the program for months
without interruption. A proprietary compression algorithm allows
real-time compression during data taking with a significantly increased
compression ratio as compared to gzip. The common framework used for
all programs keeps maintenance needs low and increases simplicity of
the individual programs. Both together ensure a stable operation of the
whole system. For remote control, a JavaScript interface is available
allowing a well controlled access to all components of the telescope
enabling the implementation of a very efficient data taking. A web
interface with low bandwidth requirements allows to control and
monitor the system easily from everywhere with just a mobile phone
under normal data taking conditions. Excluding external factors, data
taking efficiencies of more than 90\% are achieved only limited by
fixed operation times like re-positioning.

Just after data is recorded, a quick look analysis is executed
on a PC in parallel. The results of this analysis, are publicly available
online at~\cite{bib:qla}.



\end{document}